\documentclass[12pt, a4paper]{article}
\usepackage{graphicx}
\usepackage{wrapfig}
\author{V.P.Berezovoj$^\ast$, Yu.L.Bolotin$^\ast$, G.I.Ivashkevych$^\star$.}
\title{Geometrical approach for description \\of the mixed state in
multi-well potentials.}
\begin{document}
\maketitle
\begin{center}
{\footnotesize $^\ast$ National Scientific Center "Kharkov
Institute of Physics and
Technology"\\
Akademicheskaya str., 1, 61108, Kharkov, Ukraine\\ E-mail:
bolotin@kipt.kharkov.ua\\
$^\star$ Kharkov National University, Department of Physics and
Technology \\Svobody sqr., 4, 61077, Kharkov, Ukraine}
\end{center}
{\footnotesize We use so-called geometrical approach~\cite{Cerruti}
in description of transition from regular motion to chaotic in
Hamiltonian systems with potential energy surface that has several
local minima. Distinctive feature of such systems is coexistence of
different types of dynamics (regular or chaotic) in different wells
at the same energy ~\cite{bolotin1}. Mixed state reveals unique
opportunities in research of quantum manifestations of classical
stochasticity~\cite{berezovoj}. Application of traditional criteria
for transition to chaos (resonance overlap criterion, negative
curvature criterion and stochastic layer destruction criterion) is
inefficient in case of potentials with complex topology. Geometrical
approach allows considering only configuration space but not phase
space when investigating stability. Trajectories are viewed as
geodesics of configuration space equipped with suitable metric. In
this approach all information about chaos and regularity consists in
potential function. The aim of this work is to determine what
details of geometry of potential lead to chaos in Hamiltonian
systems using geometrical approach. Numerical calculations are
executed for potentials that are relevant with lowest umbilical
catastrophes.}

\section {Mixed state. Phenomenological description.}

Hamiltonian system with multi-well potential energy surface(PES)
represents a realistic model, describing the dynamics of
transition between different equilibrium states, including such
important cases as chemical reactions, nuclear fission and phase
transitions. It became known in 80-th that existence of mixed
state is an important feature of such systems~\cite{bolotin1}.
Mixed state means that there are different dynamical regimes in
different local minima at the same energy, either regular, or
chaotic. For example let's demonstrate an existence of mixed state
for nuclear quadrupole oscillations Hamiltonian. It can be shown
that using only transformation properties of the interaction the
deformation potential of surface quadrupole oscillations of nuclei
takes on the form ~\cite{mosel}:
\begin{equation} U(a_0 ,a_2 ) = \sum\limits_{m,n} {C_{mn} (a_0^2  +
2a_2^2 )^m a_0^n (6a_2^2  - a_0^2 )^n } \end{equation} where $a_0$
and $a_2$ are internal coordinates of the nuclear surface during
the quadrupole oscillations:

\begin{equation}
R(\theta ,\varphi ) = R_0 \{ 1 + a_0 Y_{2,0} (\theta ,\varphi ) + a_2 [Y_{2,2} (\theta ,\varphi ) + Y_{2, - 2} (\theta ,\varphi )]\}
\end{equation}

Constants $C_{mn}$ can be considered as phenomenological
parameters. Restricting with the members of the fourth degree in
the deformation and assuming the equality of mass parameters for
two independent directions, we get $C_{3v}$-symmetric Hamiltonian:
\begin{equation}
H = (p_x^2  + p_y^2 )/2m + U_{QO} (x,y;a,b,c)
\end{equation}
where
\begin{equation}
\begin{array}{c}
 U_{QO} (x,y;a,b,c) = \frac{a}{2}(x^2  + y^2 ) + b(x^2 y - \frac{1}{3}y^3 ) + c(x^2  + y^2 )^2  \\
 x = \sqrt 2 a_2 ,y = a_0 ,a = 2C_{10} ,b = 3C_{01} ,c = C_{20}  \\
 \end{array}
\end{equation}
Hamiltonian (3) and corresponding equations of motion depend only on
parameter $W=b^2/ac$, the unique dimensionless quantity we can build
from parameters a,b,c. The same parameter determines the geometry of
PES. Interval $0<W<16$ includes potentials with single extremum –
minimum in the origin that corresponds to spherical symmetric shape
of the nucleus. In the interval $W>16$ PES $U_{QO}$ contains seven
extremums: four minima (central, placed in the origin and three
peripheral, which correspond to deformated states of nuclei) and
three saddles, which separate peripheral minima from central one.
The distinctive feature of transition from regularity to chaos in
such a potential lies in the fact that energy of transition is not
the same in different local minima. Thus, $E_{cr} \sim E_{s}/2$
($E_s$ - energy in the saddles) for the central minimum and
$E_{cr}\sim E_s$ for peripheral. Due to this in the interval
$E_s/2<E<E_s$ classical dynamics is mainly chaotic in the central
minimum and remains regular in peripheral minima (Fig.~\ref{allp},
right). Term "mixed state" is used for designation of such specific
dynamics.

Mixed state is natural for multi-well potentials. This statement
is illustrated by Fig.~\ref{allp} (left and center), which
represents level lines and Poincare sections in different energies
for multi-well potentials from family of umbilical catastrophes
$D_5$ and $D_7$:
\begin{equation}
\begin{array}{c}
 U_{D_5 }  = 2y^2  - x^2  + xy^2  + \frac{1}{4}x^4  \\
 U_{D_7 }  = \sqrt 2 y^2  + \frac{3}{8}x^2  + xy^2  - \frac{1}{2}x^4  + \frac{1}{6}x^6  \\
 \end{array}
\end{equation}

One can see that there exists chaos in wells with three saddles, while in other wells motion is regular.
Let's note the distinction of sections structure in different wells. In the low energy there exists a hyperbolic point in the section for wells with
chaotic motion. At the same time there is no such a point in the regular wells and structure of sections is similar at the different energies.

\section {Importance of the mixed state for quantum chaos.}
\begin{wrapfigure}[13]{r}{5cm}\hspace*{1pt}
  \includegraphics[scale=0.5]{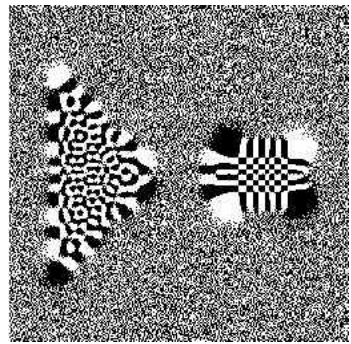}
  \caption{\footnotesize Wave function structure in $D_5$ potential}\label{waveD5}
\end{wrapfigure}

The mixed state represents optimal object for investigation of
quantum manifestations of classical stochasticity (QMCS) in wave
function structure. Indeed, usual procedure of search for QMCS in
wave functions implies distinction in its structure below and
above classical critical energy (or other parameters of
regularity-chaos transition). However, such procedure meets
difficulties connected with necessity to separate QMCS from
modifications of wave functions structure due to trivial changes
of its quantum numbers. Wave functions of the mixed state allow
finding QMCS in comparison not different eigenfunctions, but
different parts of the same eigenfunction, suited in different
regions of configuration space (different local minima of the
potential). By way of example, comparing the structure of the
eigenfunctions in central and peripheral minima of the QO
potential or in left and right minima of the $D_5$, it is evident
that nodal structure of the regular and chaotic parts is clearly
different, but correlating with the character of the classical
motion (see Fig.~\ref{waveD5}).
\begin{figure}
\begin{center}\includegraphics[scale=0.7]{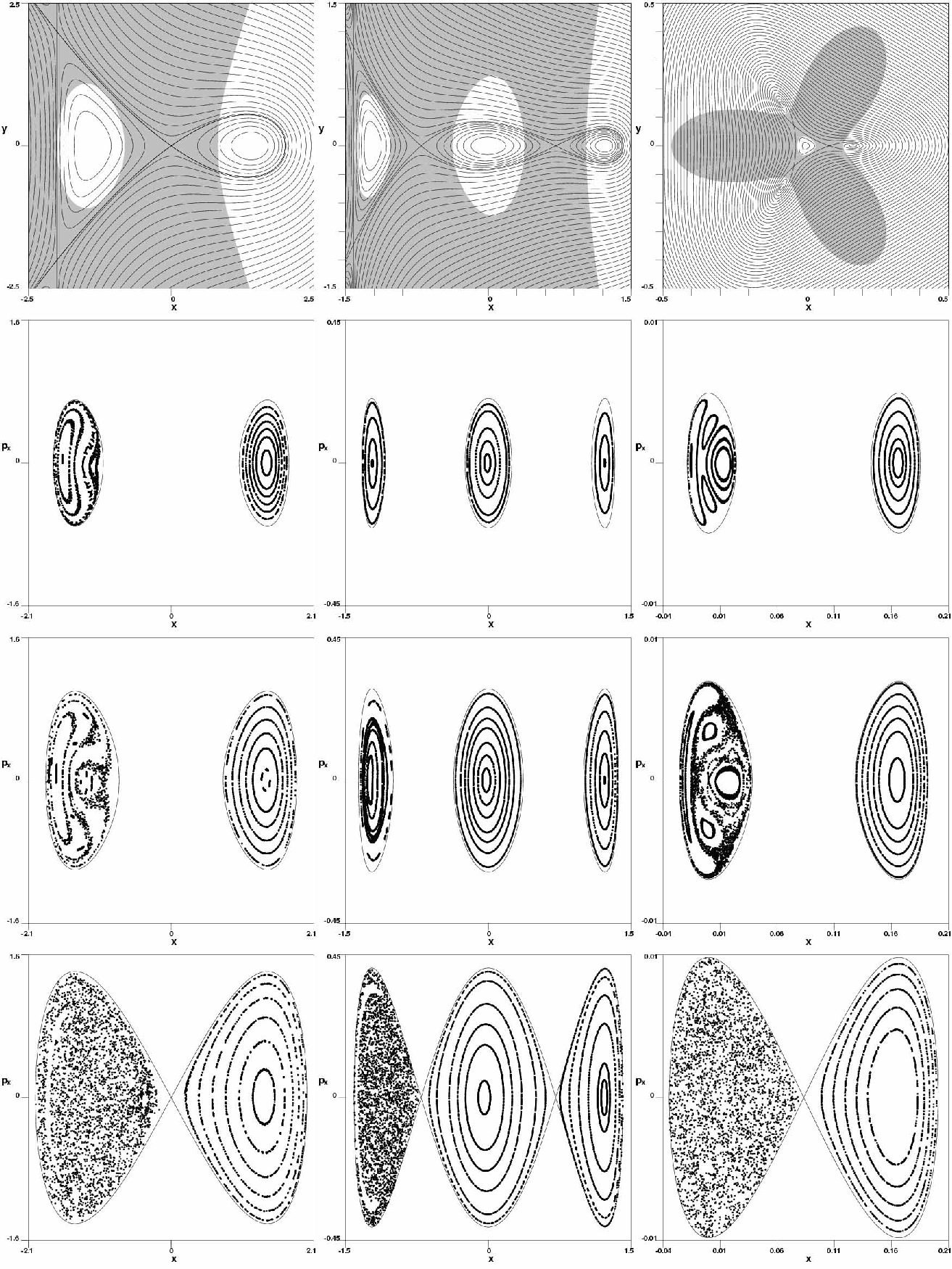}\end{center} \caption{\footnotesize
Level lines and Poincare sections for $D_5$(left), $D_7$(center) and
QO (right) potentials. Sections are presented at energies $E_s/4$,
$E_s/2$ and $E_s$.} \label{allp}
\end{figure}

\section {Stochastic criteria for mixed state.}

As well known~\cite{Zaslavsky}, stochasticity is understood as a
rise of statistical properties in purely deterministic system due
to local instability. According to this idea values of parameters
of dynamical system, under which local instability arises, are
identified as regularity-chaos transition values. However,
stochasticity criteria of such a type are not sufficient (their
necessity offer a separate and complicated question), since loss
of stability could lead to transformation of one kind of regular
motion to another one. Although this serious limitation,
stochastic criteria in combination with numerical experiments
facilitate an analysis of motion and essentially extend efficacy
of numerical calculations.

First among widely used stochasticity criteria is nonlinear
resonances overlap criterion presented by
Chirikov~\cite{Chirikov}. According to this criterion rise of
local instability is generated by contact of separatrixes of
neighboring nonlinear resonances. In this approach the scenario of
stochasticity is the following. The averaged motion of the system
in the neighborhood of the isolated nonlinear resonance on the
plane of the action-angle variables is similar to the particle
behavior in the potential well. Several resonances correspond to
several potential wells. The overlap of the resonances is
responsible for the possibility of the random walk of particle
between these wells. This method could be modified for the systems
with unique resonance~\cite{Doviel}. In this case the origin of
the large-scale stochasticity is connected with the destruction of
the stochastic layer near the separatrix of the isolated
resonance.

 Application of these criteria in presence of
strong nonlinearity (which is inevitable when considering
multi-well potentials) encounters an obstacle: action-angle
variables effectively work only in neighborhood of local minimum.
Because of this, an interest to methods, based on direct
estimation of trajectories moving away speed, arises. The
criterion of such a type is so-called negative curvature criterion
(NCC)~\cite{toda}. This criterion connects stochastisation of
motion with getting to part of configuration space, where Gaussian
curvature of PES is negative when energy increases (while in
neighborhood of minima curvature is always positive). Then energy
of transition to chaos is close to minimal energy on the
zero-curvature line. Negative curvature criterion allows getting a
number of interesting results both for classical motion and QMCS
in potentials with simple geometry – a single
minimum~\cite{berezovoj}. However, when passing on to the
multi-well potentials, NCC fails to work correct. In particular,
for above mentioned potentials ($D_5$ and $D_7$), structure of
Gaussian curvature is similar in different wells. For example, for
$D_5$ potential according to NCC we get the same value of critical
energy for both minima: $-5/9$, but chaotic motion is only
observed in the left well (Fig.~\ref{allp}, left). A natural
question immediately arises: is it possible, using only
geometrical properties of PES but not solving numerically
equations of motion, to formulate an algorithm for finding a
critical energy for single local minima in multi-well potential?
We'll try to answer on this question below in the framework of
geometrical approach.

\section {Geometrical approach to Hamiltonian mechanics.}

We will use so-called geometrical approach in consideration of
mixed state. Let's recall the basics of this
method~\cite{Cerruti}. It is known that Hamiltonian dynamics could
be formulated in the terms of Riemannian geometry. In this
approach trajectories of the system are considered as geodesics of
some manifold. Grounds for such consideration lie in variational
base of Hamiltonian mechanics. Geodesics are determined by
condition:
\begin{equation}
\delta \int\limits_L {ds = 0}
\end{equation}
At the same time trajectories of dynamical system are determined
according to the Maupertuis principle:
\begin{equation}
\delta \int\limits_\gamma  {2Tdt = 0}
\end{equation}
($\gamma$ - all isoenergetic paths connecting end points) or to the
Hamilton's principle:
\begin{equation}
\delta \int\limits_{t_1 }^{t_2 } {Ldt = 0}
\end{equation}

Once chose a suitable metric action could be rewrote as a length
of the curve on the manifold. Then trajectories will be geodesics
on this manifold(configurational - CM). This approach has an
evident advantage: potential energy function includes all
information about the system, so one need to consider only
configurational space (CS) but not phase space. Equations of
motion in this case take on the form:
\begin{equation}
\frac{{d^2 q^i }}{{ds^2 }} + \Gamma _{jk}^i \frac{{dq^j
}}{{ds}}\frac{{dq^k }}{{ds}} = 0
\end{equation}
Christoffel symbols in this approach play as counterparts of
forces in ordinary mechanics. The most natural metric is a Jacobi
one. It has the form:
\begin{equation}
g_{ij}  = 2[E - V(q)]\delta _{ij}
\end{equation}
By means of this metric Maupertuis principle could be rewrote in
the form equivalent to condition for geodesics.

Let's consider local instability in the framework of above
mentioned geometrical approach. Let  $q$  and  $q'$  be two nearby
at the $t=0$ trajectories:
\begin{equation}
q'^i (s) = q^i (s) + J^i (s)
\end{equation}
Separation vector $\overrightarrow{J}$  then satisfy the
Jacobi-Levi-Civita equation:
\begin{equation}
\frac{{d^2 J^i }}{{ds^2 }} + R_{jkl}^i \frac{{dq^j }}{{ds}}J^k
\frac{{dq^l }}{{ds}} = 0
\end{equation}

It can be shown that dynamics of the deviation is determined only by
Riemannian curvature of the manifold. For two-degrees-of-freedom
systems Riemannian curvature has a form:
\begin{equation}
R = \frac{1}{{4(E - V)^2 }}[2(E - V)\Delta V + 2\left| {\nabla V}
\right|^2 ]
\end{equation}
Laplasian of $V$ is positive for considered potentials so
Riemannian curvature is positive too. Due to this we couldn't
connect divergence of trajectories with negative Riemannian
curvature. The one way to solve this problem consists in
introduction of higher-dimensional (than $N$) metrics. Let's
examine this question closer. It can be shown that equation for
separation vector $\overrightarrow{J}$ could be reformulated in
form, which doesn't depend on dimensionality of manifold:
\begin{equation}
\frac{1}{2}\frac{{d^2 \left\| {\vec J} \right\|^2 }}{{ds^2 }} +
K^{(2)} (\vec J,\vec v)\left\| {\vec J} \right\|^2  - \left\|
{\frac{\nabla }{{ds}}\vec J} \right\|^2  = 0
\end{equation}
where $K^{(2)}$ is a sectional curvature in two-dimensional
direction:
\begin{equation}
K^{(2)} (\vec J,\vec v) = R_{iklm} \frac{{J^i }}{{\left\| {\vec J}
\right\|}}\frac{{dq^k }}{{ds}}\frac{{J^l }}{{\left\| {\vec J}
\right\|}}\frac{{dq^m }}{{ds}}
\end{equation}
and $ \left\langle {\vec J,\vec v} \right\rangle  = 0 $. Note that
point where $K^{(2)}<0$ is an unstable one. Since there are more
than one sectional curvature for the case $N>2$, we could connect
instability with negative sign of some of them. One among the
enlarged metrics is an Eisenhart metric. Eisenhart metric is
$N+2$-dimentional and contains two additional coordinates. One of
these coordinates coincides with a time and second is connected
with action. Using Eisenhart metric, quantity $K^{(2)}$ could be
rewrote in the form:
\begin{equation}
K^{(2)} (\dot{\vec {q}}  ,\vec q) = \frac{1}{{2(E -
V)}}(\frac{{\partial ^2 V}}{{\partial q_1^2 }}\mathop
{{\dot{q}_2}^2} + \frac{{\partial ^2 V}}{{\partial q_2^2 }}\mathop
{{\dot{q}_1}^2} - 2\frac{{\partial ^2 V}}{{\partial q_1
\partial q_2 }}\mathop {\dot{q}_1} \mathop {\dot{q}_{2}} )
\end{equation}

Now, investigation of the $K^{(2)}$ – structure on the considered
manifold could be used for studying of chaotic regimes and, in
particular, the mixed state.
\section {Investigation of mixed state in the framework of geometrical
approach.}
As mentioned above, negative sign of $K^{(2)}$ is a
condition for rise of local instability. It is necessary to
clarify whether this condition is sufficient for development of
chaoticity or not, clearly speaking, one needs to answer the
question, does a presence of negative curvature parts on CM always
lead to chaos. Potentials with mixed state are a very convenient
model for investigation of this question, since there exist both
regimes of motion. So, we need to study, how differs the structure
of $K^{(2)}$ in different wells. For that we calculate a part of
phase space with negative curvature as a function of energy, i.e.
a volume of phase space where $K^{(2)}<0$ referred to the total
volume:
\begin{equation}
\mu (E) = {{\int {d\vec qd\vec p  \Theta ( - K^{(2)} ) \delta
(H(\vec q,\vec p) - E)} } \mathord{\left/
 {\vphantom {{\int {d\vec qd\vec p  \Theta ( - K^{(2)} )\delta (H(\vec q,\vec p) - E)} } {\int {d\vec qd\vec p\delta (H(\vec q,\vec p) - E)} }}} \right.
 \kern-\nulldelimiterspace} {\int {d\vec qd\vec p\delta (H(\vec q,\vec p) - E)} }}
\end{equation}
An advantage of this approach consists in necessity to calculate
only geometrical properties of system without solving equations of
motion.
\begin{figure}[h]
\hspace*{1pt}
  \begin{center}\includegraphics[scale=0.9]{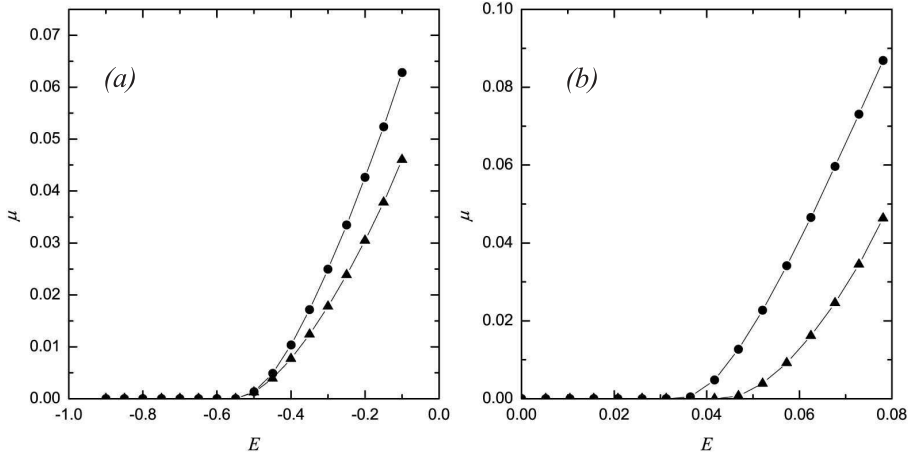}\end{center}
  \caption{\footnotesize Function $\mu(E)$ for $D_5 (a)$ and $D_7 (b)$ potentials. $\mu(E)$ for chaotic wells are represented by doted lines, for regular - by triangles.} \label{D57}
\end{figure}
\begin{figure}[h]
\hspace*{1pt}
  \begin{center}\includegraphics[scale=0.9]{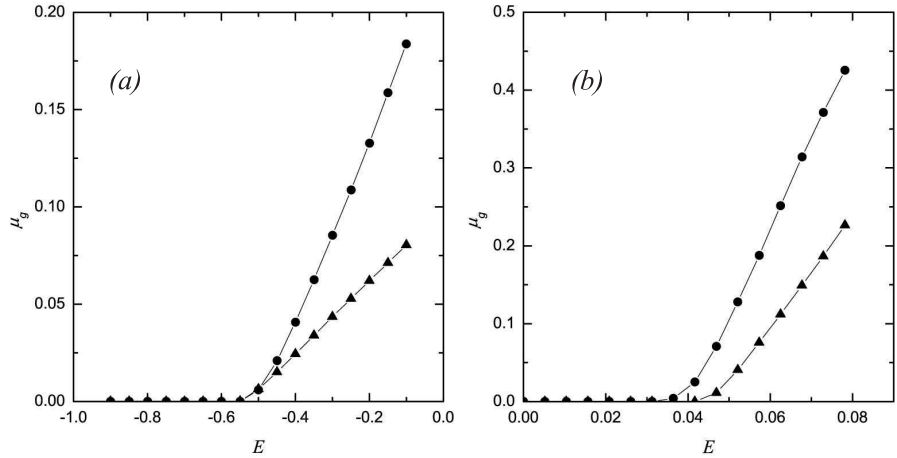}\end{center}
  \caption{\footnotesize Function $\mu_g(E)$ for $D_5 (a)$ and $D_7 (b)$ potentials.} \label{D57g}
\end{figure}

We carried out calculations for two potentials: $D_5$ and $D_7$.
Calculations of $\mu(E)$ (Figure~\ref{D57}) show that there are
parts, where $K^{(2)}<0$, in all wells, but nevertheless chaos
exists only in one well. Moreover, for well with chaotic motion
function $\mu(E)$ gives correct value of critical energy (in the
sense, specified in Section 3). In this energy $\mu$ becomes
positive. Situation with regular wells is more complex. Although
part of phase space, where $K^{(2)}<0$,  is nonzero, chaos in the
well doesn't exist. This can be viewed on the Poincare sections.
For comparison on the Fig.~\ref{D57g} part of CS with negative
Gaussian curvature is shown. One can see that structure of
negative Gaussian curvature is similar to the $K^{(2)}$-structure.

\section {Conclusions.}

Investigation of curvature of manifold, as one can see from the
cited above data, doesn't give a plain method for identification
of chaos in any minimum, especially if there exist both regular
and chaotic regimes of motion. It is impossible to determine a
priori whether chaos existed in the system without using dynamical
description (in out case that are Poincare sections).
Nevertheless, one can efficiently use geometrical methods for
investigation of chaos in multi-well potentials. In considered
above potentials chaos exists only in wells, which have two
details: non-zero part of negative curvature on the manifold and
at least one hyperbolic point in the Poincare section. According
to this, one can use the following method for identification of
chaos and calculation of critical energy. At the first step the
Poincare section in low energy is drew for the well and the
presence of hyperbolic point is determined. If so, the quantity
$\mu$ must be calculated (or the part of CS with negative Gaussian
curvature). Value of energy, in which $\mu(E)$ become positive,
could be associated with critical energy. If there is no
hyperbolic point in the section than chaos doesn't exist in the
well. Consequently, geometrical methods could be efficiently used
for determination of critical energy in complex potentials and
identification of chaos in general. However, one must carefully
use these methods and combine them with qualitative methods, such
as Poincare sectioning method.

{\footnotesize Report was made at XIII International Seminar
"Nonlinear phenomena in complex systems", Minsk, Belarus, May 16-19
2006}
\newpage
\begin{figure}
\begin{center}\includegraphics[scale=1]{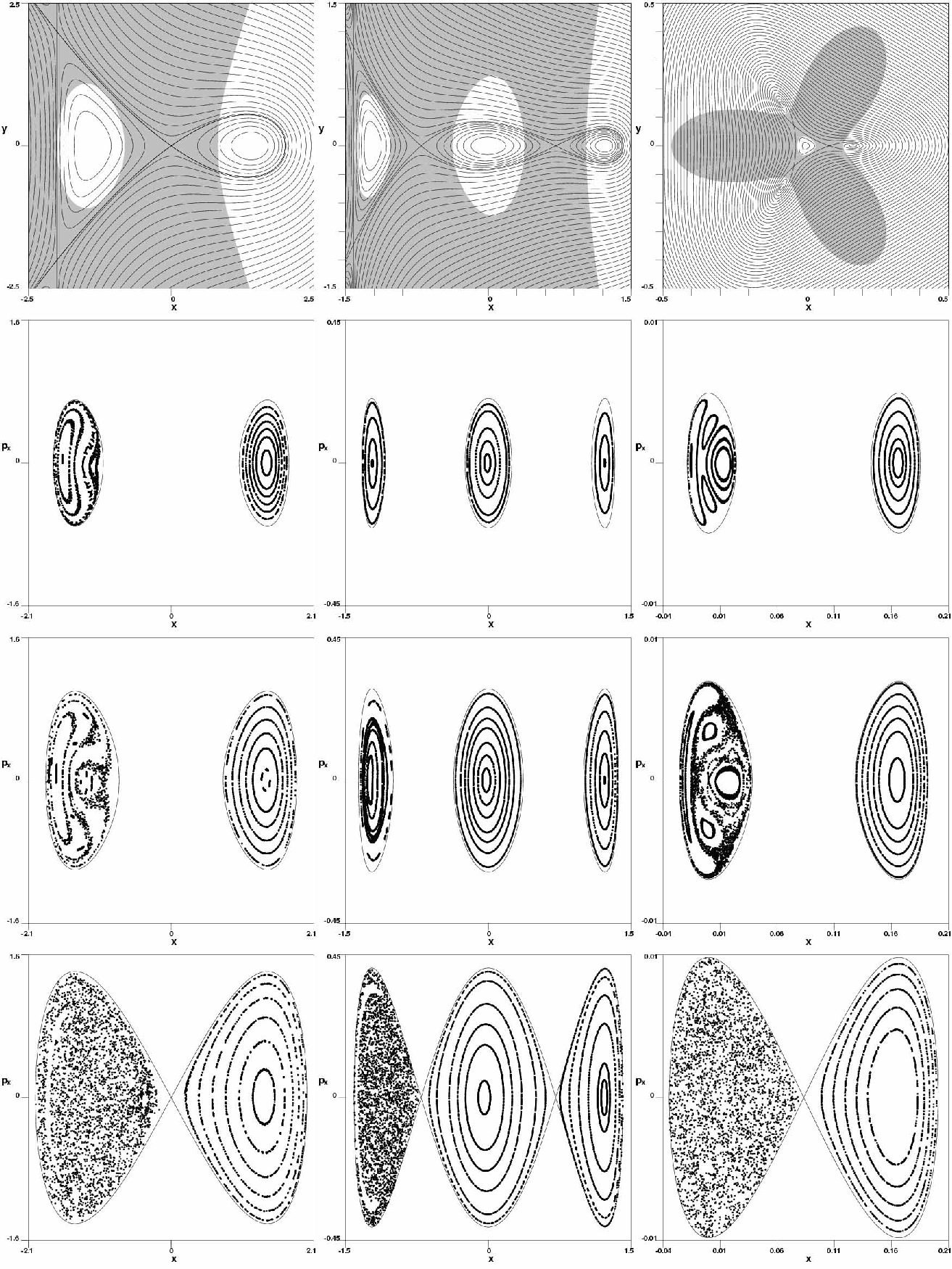}\end{center} \caption{\footnotesize
Level lines and Poincare sections for $D_5$(left), $D_7$(center) and
QO (right) potentials. Sections are presented at energies $E_s/4$,
$E_s/2$ and $E_s$.} \label{allp}
\end{figure}
\end {document}